# From Consensus to Chaos: A Vulnerability Assessment of the RAFT Algorithm

Tamer Afifi, Dr. Abdelfatah Hegazy, Dr. Ehab Abousaif
Department of Computer Science-College of Computing & Information Technology,
Arab Academy for Science, Technology & Maritime, Cairo, Egypt

*Abstract*—In recent decades, the RAFT distributed consensus algorithm has become a main pillar of the distributed systems ecosystem, ensuring data consistency and fault tolerance across multiple nodes. Although the fact that RAFT is well known for its simplicity, reliability, and efficiency, its security properties are not fully recognized, leaving implementations vulnerable to different kinds of attacks and threats, which can transform the RAFT harmony of consensus into a chaos of data inconsistency. This paper presents a systematic security analysis of the RAFT protocol, with a specific focus on its susceptibility to security threats such as message replay attacks and message forgery attacks. Examined how a malicious actor can exploit the protocol's message-passing mechanism to reintroduce old messages, disrupting the consensus process and leading to data inconsistency. The practical feasibility of these attacks is examined through simulated scenarios, and the key weaknesses in RAFT's design that enable them are identified. To address these vulnerabilities, a novel approach based on cryptography, authenticated message verification, and freshness check is proposed. This proposed solution provides a framework for enhancing the security of the RAFT implementations and guiding the development of more resilient distributed systems.

*Keywords—RAFT; consensus protocol; security; distributed systems; message forgery; replay attacks; cryptography*

## I. Introduction

The world today is surrounded by distributed systems everywhere, which operate behind the scenes, seamlessly coordinating tasks across multiple computer-based systems, often geographically scattered [2]. From global communication networks to critical financial transactions, these systems rely on fundamental principles of distributed systems to ensure reliability and data consistency [1]. A core component of this reliability is the distributed consensus algorithm, which enables autonomous nodes to agree on a single value, even in the presence of network failures or node crashes [9]. In recent decades, the RAFT distributed consensus algorithm has become a main pillar of the distributed systems ecosystem.

The RAFT algorithm is a foundational element for ensuring data integrity in distributed consensus. Designed for understandability and implementation ease [3], it has not only simplified the process of building resilient distributed applications but has also become a standard of comparison for new consensus research. RAFT achieves its primary goal of fault tolerance by replicating state across nodes, making it resilient against system failures and crashes.

However, while RAFT is well-known for its fault tolerance and simplicity, the consensus mechanism itself typically operates under a "fail-stop" model (node crashes or network partitions), assuming non-malicious failures [10]. Consequently, the security concerns regarding its design, particularly its mitigation against active, malicious attacks, are often not fully integrated or covered in standard implementations [4, 5]. This gap leaves implementations vulnerable to threats and attacks that can compromise the RAFT state machine and data integrity established by the protocol, potentially leading to data corruption and service disruption in critical applications [8].

This gap in the existing RAFT implementation poses a significant risk of vulnerability; a secure consensus algorithm is more than just ensuring agreement; it's about safeguarding the entire ecosystem of distributed systems, protecting data, preventing disruptions, and ultimately, mitigating failures and data corruption. In particular, the fact that the protocol does not have its own end-to-end security mechanisms makes it vulnerable to potential threats, including replay attacks and message forgery attacks, which target the RAFT's message passing mechanism and disrupt the consensus process. Therefore, this paper, in the coming sections, analyzes and evaluates the RAFT algorithm's security properties, demonstrating its design weaknesses with a focus on its susceptibility to message replay attacks and forgery attacks. The primary objective of this research is to propose and validate a novel, modular, and lightweight security enhancement that is based on encryption and authenticated message verification to mitigate these vulnerabilities and enhance the RAFT security implementations.

To achieve this objective, the subsequent sections of this paper first build the necessary foundational Distributed Systems and Consensus Background in Section II. Then Section III dives deeper into the details of the RAFT Distributed Consensus Protocol. Section IV provides a thorough Security Analysis of the RAFT vulnerabilities, focusing on message integrity. Section V presents the Related Work concerning secure RAFT implementations and countermeasures, justifying the research gap. Section VI introduces the Proposed Solution for mitigating the identified security concerns. Section VII details the Methodology used for testing and validation, followed by Section VIII, which presents the Results of the simulated attacks and the performance evaluation. Section IX provides a dedicated Discussion of the findings, their implications, and researchers point of view. Section X offers the Conclusion of this research outcome. Finally, Section XI outlines directions for future study.

## II. Background

### A. Distributed Systems

Distributed systems are the core basis of today's technology ecosystem, from powering global communication networks to orchestrating critical financial transactions, distributed systems can be defined as a collection of autonomous computing elements (nodes) that appears to users as a single coherent system [1], within this single system the collection of nodes regardless of their number, locations, or components, operate as a unified whole, no matter where, when, and how interaction between a user and the system takes place.

The distributed system provides the means for components of a single distributed application to communicate with each other. At the same time, it hides the differences in hardware and operating systems from each application; this can be orchestrated using a middleware.

In a sense, middleware to a distributed system is the same as an operating system to a computer: a manager of resources offering its applications to efficiently share and deploy those resources across a network [2].

The basic distributed systems architecture and the relationship between its main components: hardware, applications, the operating system, and the middleware layer, which acts as a very essential role in coordinating communication and abstracting hardware differences across the network, is illustrated in Fig. 1.

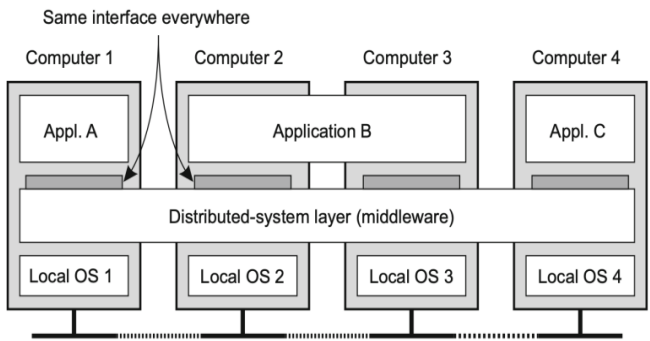

Fig. 1. The middleware layer extends over multiple machines and offers each application the same interface.

The middleware, nodes, applications, and this ever-expanding realm of distributed systems need to assure stability and consistency of communication and exchanged messages, and keep all parties informed of all correct and recent updates on the system, here the important role of the consensus algorithm, which addresses these needs, is thus witnessed.

### B. Distributed Consensus

Consensus in general means agreement made by multiple parties; for example, if a group of friends decides to have lunch, which restaurant to order from is an agreement. Basically, consensus has the same meaning in computer science, especially in distributed systems, as it can be defined as a process where multiple nodes of a distributed system agree on values of messages, transactions, or objects. It is a basic challenge in distributed systems. However, once the nodes agree on a value, that agreement should be final.

Some of the earliest implementations of consensus algorithms relied on voting-based mechanisms, which provide reasonable fault tolerance and have strong mathematical proofs to ensure integrity and stability. Some of the popular voting-based algorithms include Paxos and Raft. Paxos was originally first proposed by Leslie Lamport in 1989, however it was published by the end of 2001 due to many factors, including, of course, its well-known high level of complexity, which hindered its widespread adoption. In response to this, the Raft algorithm was introduced in 2014 by Diego Ongaro as a simpler alternative to Paxos. Raft's design emphasizes simplicity and manageability while maintaining the same level of fault tolerance and consistency as its predecessors.

## III. RAFT Distributed Consensus Protocol

This section provides a brief overview of the Raft distributed consensus algorithm. For a more detailed description, please refer to the original paper "In search of an understandable consensus algorithm" [3].

### A. The Raft

Raft is a highly common and reliable distributed consensus algorithm designed as a more understandable and easily implementable alternative to its complex predecessor, the Paxos algorithm. Raft is a fault-tolerant protocol that depends on a single elected leader, log replication, and a probability approach.

To understand how Raft works, let's imagine a server cluster of three nodes or replicas, each hosts a state machine, log, and raft protocol, as long as they all begin with the same state and perform the same operations in the same order then they should all end up with the same state, anytime a replica receives a command such as setting a new key with a value, the replica appends and saves the command as a new entry in its log, every replica's log must always contain the same exact sequence of commands for other replicas to remain synchronized.

In a Raft cluster, one node is elected as the leader, with the others acting as followers. The leader is responsible for handling all client requests and for replicating log entries to the followers to ensure they remain synchronized.

### B. Leader Election

At any point in time, any node of the Raft cluster can take on only one state: leader, follower, or candidate. Initially, all nodes start in the follower state. Each follower sets an election random timeout. Thus, if a follower fails to receive heartbeats consecutively, then it assumes there is no viable leader and transitions to candidate state to start an election [13]. It wins the election and becomes the new leader if it secures a majority of the total votes. Once elected, a node remains a leader until it crashes or observes another node with a higher term [15]. If the election results in a split vote or if another candidate is elected, the node reverts to the follower state to await a new leader's heartbeat. The RAFT leader election flow is illustrated in Fig. 2.

Leader election can begin for any reason, whether it's brought about after a leader fails, goes offline, if the network experiences enough latency, or when a network partition isolates a follower where a follower reaches its election timeout despite a leader still being alive.

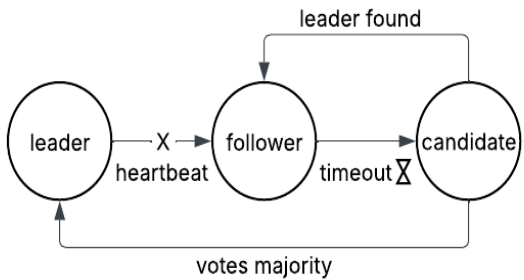

Fig. 2. The Raft basic leader election flow.

Once a leader is elected, it assumes responsibility for logging all new changes to the system. The leader regularly sends append entry messages to all followers within the RAFT cluster. These messages serve a dual purpose: they act as a heartbeat to prevent followers from initiating a new election, and they instruct followers to replicate new log entries, ensuring consistency across the cluster.

*C. Log Replication*

All changes in the system, such as new commands or transactions, are exclusively handled by the leader. Changes are sent by clients to the leader who receives them and appends a new entry to its log, this log entry remains uncommitted till the leader replicates it to all follower nodes, then the leader waits until a majority of nodes in the cluster (>50%) confirms/acknowledges the new value, now the leader commits the new entry, and informs the followers that the entry is committed, and the cluster is considered to have reached consensus. Fig. 3 shows the basic log replication flow of the RAFT.

According to Ongaro and Ousterhout [3], founders of the RAFT consensus [19, 21] algorithm, it was originally designed to be more understandable than Paxos, simpler, and more efficient, which later became a foundation for a wide range of applications requiring fault-tolerant data storage and consistent state management. However, despite its resilience and robustness in handling system failures, the RAFT's message-passing architecture remains vulnerable to various security threats that need to be investigated further, including message forgery and replay attacks.

IV. RAFT SECURITY ANALYSIS

Although that Raft consensus algorithm is a key technology for state replication in distributed systems [14] and is well-known for its simplicity, effectiveness, and dependability, its security vulnerabilities remain a critical area of exploration to consider. This section investigates some of the known threats, attack methods, and discusses the potential consequences of compromising consensus and jeopardizing system stability.

*A. Denial-of-Service (DoS) Attacks*

A denial-of-service (DoS) attack is a malicious attempt by an attacker to overload the leader or other nodes with a flood of messages, preventing them from operating normally and overloading their resources.

Since Raft adopts a strong leader model, a malicious leader can launch DoS or censorship attacks by intentionally delaying or dropping messages, which disrupts log replication and causes followers to initiate unnecessary elections. Moreover, an attacker can control multiple nodes and, through a majority vote, ensure that one of their compromised nodes is elected as the leader [4].

*B. Byzantine Attacks / Message Forgery or Impersonation*

In the RAFT consensus algorithm, the system's integrity can be compromised by Byzantine attacks, which include message forgery and impersonation, launched by nearby malicious/illegitimate node(s). Indeed, additions can be seen as messages surreptitiously inserted in the system by some outside, and possibly malicious, entity [12]. In the impersonation attacks, malicious nodes try to claim themselves as legitimate nodes/followers by utilizing a forged character in order to destroy the consensus mechanism [5].

Byzantine attacks sending fake messages may result in transaction latency, data inconsistencies, data leaks, and even compromising system integrity.

This can have a measurable impact on performance. For example, transaction latency can be modeled to show its linear relationship to the number of attackers. The total commit time ($t_c$) is the sum of replication time ($t_r$) and transaction latency ($t_l$):

Transaction commit time can be calculated as follows:

$$t_c = t_r + t_l$$

And transaction latency as

$$t_l = RTT(1 + q)(1 + p)$$

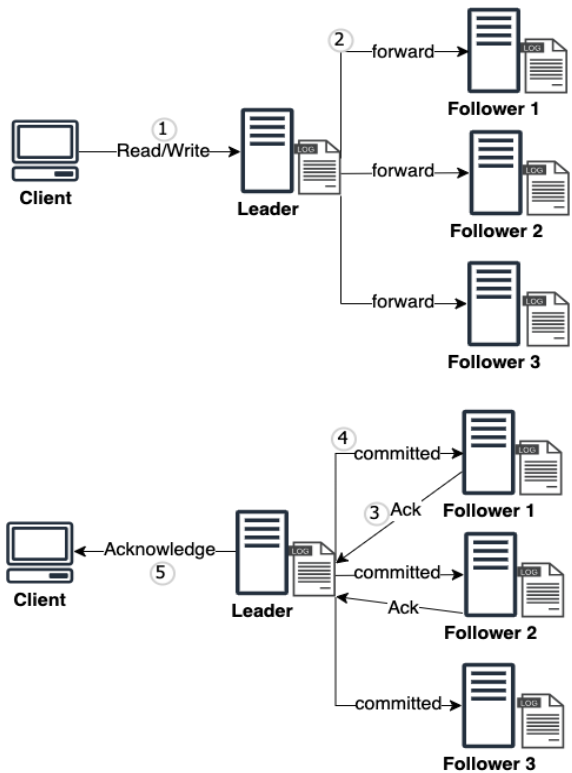

Fig. 3. The Raft basic log replication flow.

This shows that if the number of attackers increases, the transaction latency would linearly increase as well.

TABLE I. NOTATIONS

| Notations | Description |
|---|---|
| $t_c$ | transaction commit time |
| $t_r$ | transaction request time |
| $t_l$ | transaction latency |
| RTT | round-trip time for request/response message |
| $p$ | percentage of Byzantine nodes (= n/m) |
| $q$ | attack success rate |
| $n$ | number of Byzantine nodes |
| $m$ | number of all nodes |

### C. Message Replay Attacks

The previous general vulnerabilities set the grounds for more complex attacks. While many types of attacks represent notable threats, the message replay attacks represent a serious and often overlooked threat to consensus protocols like RAFT. A message replay attack is a type of network attack in which a malicious actor intercepts a legitimate data transmission, captures a valid message, and maliciously retransmits it later. In the context of distributed consensus, this attack takes advantage of the protocol's reliance on message-passing to deceive nodes and disrupt the consensus process and the system's state (see Table I for notations).

In the RAFT cluster, an attacker can start a replay attack by capturing valid, previously sent messages like RequestVote and re-injecting them into the network again later, which can cause a follower or more to respond with a vote, consequently triggering a new, unnecessary election and causing a leader-follower split.

The success of a message replay attack on RAFT is primarily due to the protocol's vulnerability to a lack of intrinsic replay protection mechanisms. Standard RAFT messages do not include a monotonically increasing sequence number or any unique, session-based identifier that would allow a recipient node to detect that a message is a stale duplicate. This absence of a freshness check makes the protocol susceptible to being manipulated by replayed messages, leading to a breakdown in consensus. This can result in system instability, data inconsistencies, and a compromise of the distributed system's integrity.

## V. RELATED WORK

The increasing dependency on distributed systems for critical infrastructure/systems has highlighted the need for robust consensus mechanisms that are resilient to both crash failures and malicious attacks. In this section, relevant researches are reviewed to compare and validate this paper's proposed solution for mitigating replay/message forgery attacks against the default RAFT implementation, and other RAFT flavors that addressed these concerns.

The theoretical foundation of achieving fault tolerance in distributed systems is rooted in the State Machine Replication approach [9], which requires all replicas to execute the same sequence of operations in the same starting state [1]. To achieve this goal, we have to make sure no malicious factor is affecting the state machine and the exchanged messages integrity. The RAFT consensus protocol [3] was introduced as a practical alternative to Paxos, focusing on understandability and simplicity while achieving the same fault-tolerance properties (tolerating crash failures). Standard RAFT implementations, however, assume a "fail-stop" environment and do not account for the Byzantine Faults, where nodes can behave maliciously and affect the integrity of the consensus process. This distinction is crucial, as general Byzantine protocols like P-BFT [20] often introduce high computational overhead trying to address this problem, which, on the other hand, prompted research into lightweight alternatives that can keep the performance benchmarks while achieving the security mitigations effectively.

The standard RAFT design assumes a "trusted" environment where nodes are honest but can crash. This assumption left the protocol inherently vulnerable to active attacks that exploit its unauthenticated, plaintext message structure. Several studies have tried to identify, address, and formalize these weaknesses:

*1) The Original RAFT:* The foundational work by Ongaro and Ousterhout [3] introduced RAFT as a more understandable alternative to Paxos, focusing on maintaining consistency against crash failures and network partitions, implicitly assuming a trusted network environment, however they did not consider security attacks such as replay attacks and message forgery.

*2) RaBFT:* an improved Byzantine fault tolerance consensus algorithm based on Raft [10]. Full Byzantine Fault Tolerance (tolerate f malicious nodes in 3f+1 total nodes), where it introduces Secret Sharing to optimize log message distribution and uses a dynamic Committee role to distribute leader pressure. It alters the log replication process and election logic. It protects against malicious nodes forging; however, it requires a complete redesign of the RAFT consensus engine, introducing complex multi-party computation steps (secret sharing verification) and new roles (Committee), and adds loads of overhead layers burden.

*3) ENGRAFT (Enclave-Guarded RAFT)* is another example of a security enhancement that operates within the consensus layer, specifically designed to protect against Byzantine faults within the nodes themselves [4]. It leverages Trusted Execution Environments (TEEs), such as Intel SGX enclaves, to safeguard the core RAFT state and logic on each node. In ENGRAFT, all critical consensus operations, including state updates and log manipulation, are performed inside the secure, isolated hardware enclave. This design is highly effective at defending against insider threats and BFT attacks, however this approach introduces dependencies on specific hardware and results in a higher implementation complexity in terms of hardware, protocol, and platform.

*4) Countering Active Attacks on RAFT-Based IoT Blockchain Networks [5] using pathloss:* This paper proposes a physical-layer authentication mechanism using pathloss to secure the Raft against impersonation attacks in a wireless IoT environment. The proposed solution uses pathloss of the signal between the transmitter node and the receiver node as a unique device fingerprint to authenticate the sender. This solution's weak points are that it can be easily affected by environmental changes and multipaths.

*5) Trust and Reputation Management System [11]:* This method assigns credit scores to nodes based on historical behavior to identify and isolate malicious nodes. However, such systems suffer from the "cold start" problem and slow convergence; malicious nodes can behave honestly for long periods to build high trust. Furthermore, the overhead of storing and updating reputation scores for every node scales poorly in large distributed systems.

*6) Combination of P-BFT and RAFT:* "A New Approach to Building Networks that Provide Reliability and Security" [20] proposes a hybrid model that merges the simplicity and efficiency of RAFT with the robust security of P-BFT. This combination aims to build networks capable of withstanding Byzantine failures (malicious behavior). While highly effective at securing the consensus process against internal attacks, BFT-RAFT inherently requires significant modifications to the RAFT state machine, leading to increased complexity in protocol execution and a higher overhead.

*7) Zero Trust Consensus Algorithm [16]:* This notable effort to bridge the gap between RAFT and modern security concepts proposes a VSSB-Raft, which adopts a Zero Trust security model, "never trust, always verify". It achieves this high security by leveraging verifiable secret sharing (VSS) mechanisms and digital signatures to ensure that no single node can compromise the system. While VSSB-Raft demonstrates a solution for achieving Byzantine Fault Tolerance in a resource-efficient manner compared to traditional BFT protocols, its comprehensive security guarantees rely on significant modifications to the RAFT state machine and the integration of complex cryptographic primitives, leading to a higher implementation complexity than modular transport layer solutions.

*8) GPBFT (Group-Based Practical Byzantine Fault-Tolerant)[17]:* an enhancement to the existing BFT protocols, involves leveraging a Dual Administrator Short Group Signature mechanism. Allowing nodes to verify messages using short group signatures, while powerful, these solutions maintain a high barrier to entry, adding significant complexity and communication overhead, plus it fully replaces the RAFT with a complex BFT protocol.

*9) PB-Raft [18]:* integrate Byzantine Fault Tolerance (BFT) capabilities into RAFT. This solution is dual-layered: first, introduces a Weighted PageRank algorithm to evaluate the reputation and trust of nodes. Second, utilizes a BLS (Boneh–Lynn–Shacham) threshold signature scheme to ensure the authenticity and consensus of log entries. This solution introduces two main points of complexity and potential overhead plus that the leader must collect a threshold of signatures before committing a log in addition to that the reputation scoring based on PageRank requires continuous, complex calculation and updates across the cluster.

The review of related work reveals a clear trade-off within the current landscape of RAFT security. While some of the previous work in this field is mature enough, offering comprehensive security against malicious attacks, they incur significant overhead and require complex modifications to the core consensus logic. This analysis confirms a significant gap in the literature: the lack of a modular, low-overhead security solution that operates at the transport layer to specifically and comprehensively counter active message forgery and replay attacks on the fundamental RAFT without altering the core of the consensus protocol itself. Therefore, the objective of this work is to introduce a secure transport layer mechanism that fills this gap, providing essential message authentication and freshness without altering the RAFT state machine, as detailed in the following section.

## VI. PROPOSED SOLUTION

As discussed in this paper, the Raft distributed consensus protocol, its mechanism, and its importance, also explored some of its security risks.

It is intended in this section to mitigate the vulnerabilities to message replay attacks and propose a secure transport layer integration to the RAFT, designed to protect communication between cluster nodes in a modular way, leaving the original RAFT intact. It uses a combination of modern cryptographic techniques to ensure message confidentiality, authenticity, some sort of checksum for integrity, and a message cache for replay attacks mitigation. This proposed solution is meant to be a lightweight modification to the original RAFT algorithm, ensuring that it remains efficient while significantly boosting its security properties. The proposed solution architecture is shown in Fig. 4.

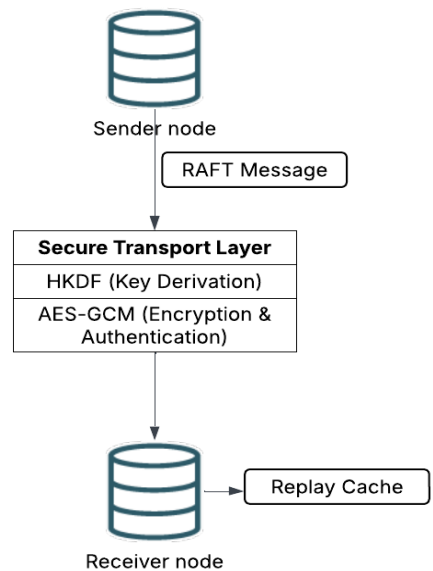

Fig. 4. Proposed solution basic flow.

The Proposed approach involves three key components:

*A. Unique Encryption Key*

The proposed solution uses the HMAC-based Key Derivation Function (HKDF) [6] to generate a unique, temporary encryption key for every single message, rather than depending on a single, long-term key for all communications. HKDF derives a fresh key for each message from a shared master secret. This procedure is a very important security best practice, because if an attacker were able to compromise one message's key, they would not be able to compromise any other messages, and the revealed key would be useless.

*B. Authenticated Encryption*

The AES-GCM Advanced Encryption Standard (AES) in Galois Counter Mode (GCM) [7] is the cryptographic algorithm used for both encryption and authentication of RAFT messages during communication in the proposed solution.

- AES: This part of the algorithm ensures that the message content is encrypted, so no one can read it in transit. To guarantee the most complex level of encryption in the AES process, the GCM [7] mode will be used.
- GCM: This part generates an authentication tag that guarantees the message has not been tampered with and that it originates from a valid sender with the correct key. This is vital in RAFT, where nodes must trust that a log entry or vote is legitimate and unaltered.

Below is a pseudo-code of the algorithm that explains it:

function secure-envelope(plaintext, confidential, peer_id, K_master, key_id):

// generate unique per-message identifiers

nonce ← random(12)

tx_id ← random(16)

// derive a unique key for this message

K_tx ← HKDF-SHA256(K_master, nonce, peer_id)

// construct Associated Authenticated Data (AAD)

aad ← combine(key_id, nonce, tx_id, peer_id)

if confidential:

// encrypt and authenticate with AES-GCM

ct_combined ← AESGCM.encrypt(K_tx, nonce, plaintext, aad)

ciphertext ← ct_combined[:-16]

tag ← ct_combined[-16:]

else:

// authenticate only (generate tag)

ciphertext ← plaintext

tag ← compute_tag(K_tx, aad, plaintext)

end if

// return the final secure envelope

return { key_id, nonce, tx_id, ciphertext, tag}

end function

The AES-GCM algorithm architecture and full flow, starting from the initialization vector and till generating the tag which is attached to the encrypted message, as illustrated in Fig. 5.

*C. Replay Cache*

A replay cache is a critical additional defense layer against replay attacks. The replay cache works by maintaining a list of identifiers (the unique message ID) for all recently processed messages. When a new message arrives, the system checks its identifier against the cache. If the ID is found, the message is immediately rejected as a replay, preventing the attack from succeeding, else the message is accepted, and its ID is logged in the replay cache.

Below is a pseudo-code explaining the cache check:

data_structure: cache # internal data structure// Function to check if a message has been seen before function SEEN(peer_id, tx_id): key ← combine(peer_id, tx_id)   if key in cache:       return True // Replay detected    else:    return False // New message    end if end function   // function to record a new message as seen function REMEMBER(peer_id, tx_id):key ← combine(peer_id, tx_id)   cache.add(key) end function.

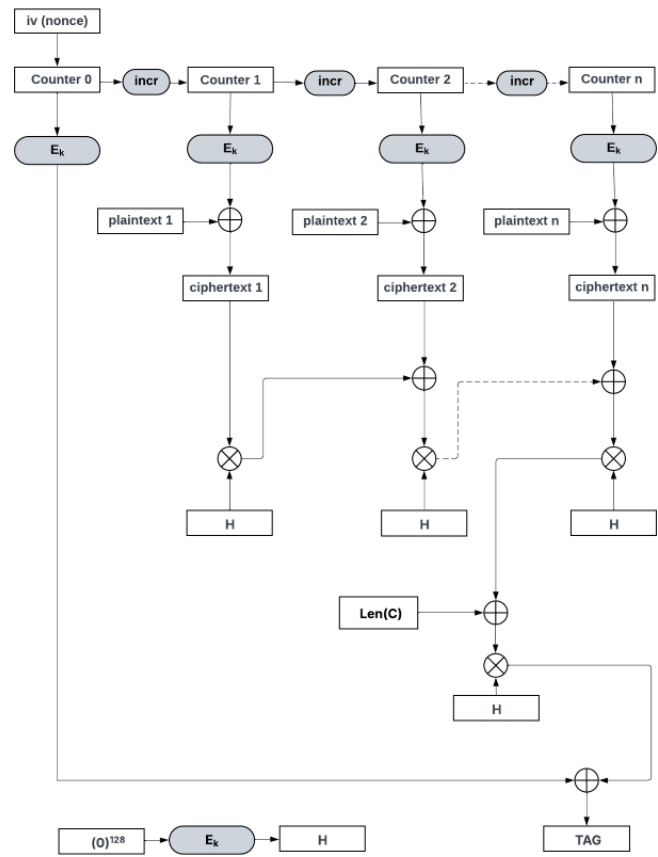

Fig. 5. AES-GCM.

## VII. METHODOLOGY

This section provides an overview of the methods and tools used to evaluate the proposed solution for mitigating replay attacks and message forgery on a RAFT cluster. The primary objective is to evaluate the impact of a replay attack on a RAFT cluster before and after the implementation of the proposed security solution to prove the RAFT's vulnerability and the efficiency of the proposed solution.

*A. Experimental Setup and Architecture*

The study's methodology and architecture focuses on a python based native RAFT implementation based on https://github.com/nikwl/raft-lite, a physical lab setup of three nodes in a RAFT cluster, all of the three nodes are running the Linux (Ubuntu server) as an operating system, and they communicate through Ethernet network using IPv4 addresses, the attack simulator and proposed solution are both developed using python as shown on Fig. 6.

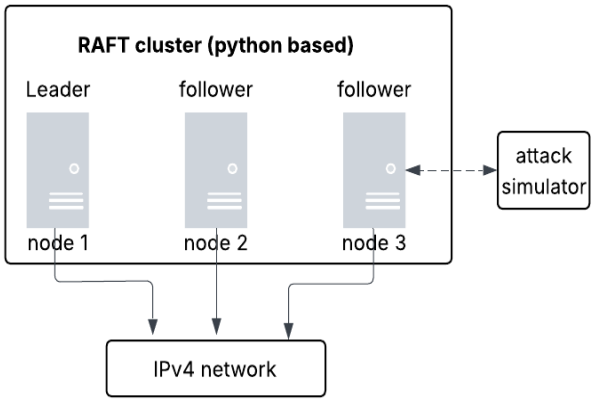

Fig. 6. The physical lab architecture.

*B. Testing Procedures*

The testing process was divided into two distinct phases to evaluate both the baseline vulnerability of the standard default RAFT behaviour before implementing the proposed solution and the efficacy of the proposed solution after implementing it.

*1) Pre-implementation testing (baseline vulnerability):* The initial phase focused on subjecting the default RAFT cluster to a simulated message replay attack to establish a security baseline we can measure based on.

*a) Attack simulation setup:* A custom client script, utilizing Python's socket library to establish a TCP connection to a follower node of the three-node RAFT cluster.

*b) Attack execution:* A valid RAFT message (e.g., a heartbeat or log entry) was captured and subsequently replayed with no modifications using a specialized attack script (replaylast) and (entryattack) to simulate log replication to introduce malicious or stale log entries, representing an explicit Byzantine message forgery and replay attack scenario.

*2) Post-implementation testing (proposed solution efficacy):* The second phase involved implementing the proposed security mechanism across all three nodes of the RAFT cluster and enabling the secure features. The same attack simulation is then activated again, and the cluster's behavior is monitored, same as done in phase one.

*a) Testing procedure:* The exact same steps and attack scripts used in the pre-implementation phase were executed against the modified cluster after activating the secure features.

*b) Solution results:* The modified RAFT implementation of the proposed solution successfully rejected and dropped all attack messages for the following reasons:

- Legitimate Traffic: Throughout the testing, all legitimate traffic, current messages continued to be accepted and processed correctly, confirming the solution's targeted function.
- Replayed Legitimate Messages: Legitimate messages that were captured and replayed later on were dropped after the receiver checked and found their unique ID already present in the local cache (failing the freshness check).
- New Forged Messages: Newly forged malicious messages were dropped due to a failure in the integrated authentication check against the tag introduced and explained earlier.

*3) Performance evaluation:* Following the proposed solution security enhancements, testing and validation, performance tests were conducted to evaluate the overhead that is introduced by the proposed solution compared to the default RAFT implementation. Key performance benchmarks were measured using the industry standard tool raft-bench in these tests.

## VIII. RESULTS

This section presents the objective findings from the security and performance tests without detailed interpretation, which will be reserved for the Discussion section.

*A. Security Assessment Results*

The security assessment confirmed the vulnerability of the default RAFT implementation and the effectiveness of the proposed solution:

*1) Pre-implementation testing (baseline vulnerability):* As expected, the default RAFT implementation lacks basic replay protection and, therefore, successfully accepted and processed the attack messages. This action disrupted the log consistency and eventually compromised the consensus state.

*2) Post-implementation testing (proposed solution efficacy):* The post-implementation testing phase demonstrated the effectiveness of the proposed Secure Transport Layer. As the modified RAFT implementation of the proposed solution successfully rejected and dropped all attack messages for the following reasons:

*a) Legitimate traffic:* Throughout the testing, all legitimate traffic, current messages continued to be accepted and processed correctly, confirming the solution's targeted function.

*b) Replayed legitimate messages:* Legitimate messages that were captured and replayed later on were dropped after the receiver checked and found their unique ID already present in the local cache (failing the freshness check).

*c) New forged messages:* Newly forged malicious messages were dropped due to a failure in the integrated

authentication check against the tag introduced and explained earlier.

*3) Performance benchmarks:* The performance tests evaluated the overhead of the proposed solution compared to the default RAFT implementation. The results are summarized in Table II.

TABLE II. PERFORMANCE ANALYSIS

| Metric | Default Raft | Proposed Solution | Change % |
|---|---|---|---|
| *Throughput* | 297.19 | 269.61 | -9.28% |
| *Latency* | 468.01 | 539.44 | 15.26% |

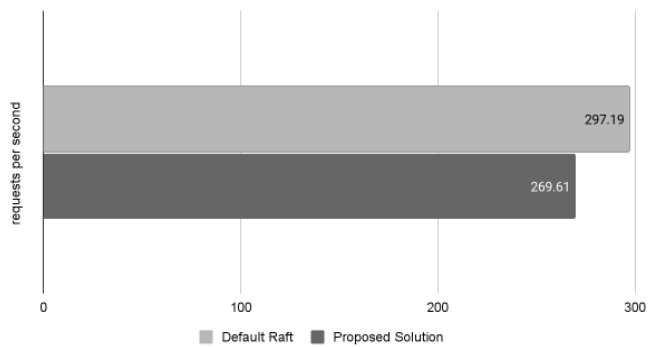

Fig. 7. Throughput performance analysis for default vs. proposed solution.

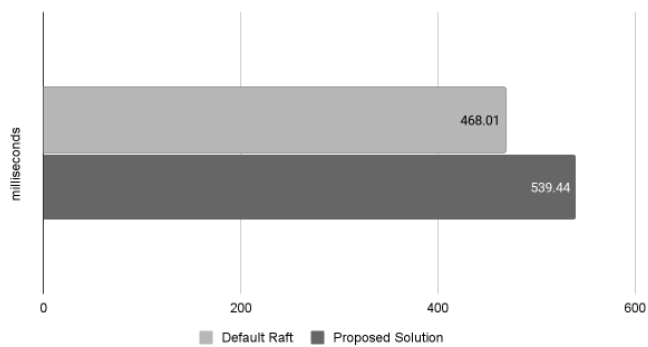

Fig. 8. Latency performance analysis for default vs. proposed solution.

As illustrated in Fig. 7, 8, and Table II, the results show a 9.28% decrease in throughput and a 15.26% increase in latency for the proposed solution. This performance cost is directly attributed to the computational overhead of generating and verifying the secure, encrypted messages and unique identifiers on every message, demonstrating a typical security-performance trade-off.

## IX. DISCUSSION

The findings of this research demonstrate that the proposed solution can effectively mitigate the two critical attacks on the RAFT: message replay and message forgery. The pre-implementation baseline testing confirmed that the standard RAFT design is vulnerable to message replay and forgery attacks due to its reliance on unauthenticated messages and the lack of a freshness check.

The post-implementation analysis showed that the proposed solution successfully rejected all replayed legitimate messages and all newly forged malicious messages. These efficient outcomes result from a combination of cryptographically enforced message authentication and the transaction ID-based replay cache. The authenticated encryption (AES-GCM) ensures the legitimacy of the sender and the integrity of the message content, while the unique, per-message identifier used by the Replay Cache successfully guarantees timing-based replay attacks mitigation.

Security-Performance Trade-off: The performance evaluation revealed a typical, quantifiable trade-off between security and efficiency. The proposed solution introduced a 9.28% decrease in throughput and a 15.26% increase in transaction latency when compared to the default, unsecured RAFT implementation. This performance cost is directly attributed to the computational overhead required for generating a unique encryption key (HKDF) and performing authenticated encryption (AES-GCM) for every message sent and received.

The key insight of this research is that the security gained—protection against system-compromising attacks like forgery and replay—comes at a price tag of this minimal overhead, especially in mission-critical environments such as financial systems or sensitive data ledgers, where data integrity is crucial. This measured performance impact is an acceptable cost for achieving a high level of data integrity and authenticity.

Comparison with Alternative Approaches: The modular design of the solution is a key differentiator. Unlike other advanced security proposals like RaBFT [10], which requires a complete redesign of the core RAFT consensus engine, or pathloss-based solutions [5], which rely on specialized physical-layer measurements, our proposal operates entirely at the transport layer. This approach achieves its primary goal of mitigating replay and forgery attacks with low complexity and minimal impact on the core RAFT logic. Which also enhances the development lifecycle of each component without affecting the other components or even replacing a whole module with a better-performing one in the future.

Limitations: The primary operational limitation of this approach is the potential for replay cache size growth over long-term operation. This needs to be managed in the future to prevent excessive memory consumption.

## X. CONCLUSION

This Paper successfully assessed the security vulnerabilities of the RAFT distributed consensus protocol, identifying its critical vulnerabilities to active message replay and forgery attacks. To mitigate these threats, a novel and modular Secure Transport Layer integration based on authenticated message verification and a freshness check was proposed. This paper's contribution, unlike previous work that either redesigns the core RAFT protocol, adds heavy burdens, targets only a specific message vulnerability (like message forgery), or relies on specialized hardware/physical-layer attributes, a modular Secure Transport Layer approach is proposed without altering the core consensus logic or adding notable overheads.

The scientific value added by this work is underscored by the following contributions:

*1) Vulnerability validation:* Practical simulation demonstrating the ease of disrupting the consensus and compromising the data integrity of the default RAFT protocol, with which unauthenticated RAFT messages can be exploited to achieve such.

*2) Modular solution:* The development and validation of a lightweight, Transport-Layer-based security enhancement that utilizes HKDF [6] for unique key derivation and AES-GCM [7] for simultaneous encryption and authentication can be achieved without changing the default RAFT logic or adding heavy burdens.

*3) Replay attacks protection:* The introduction of a unique transaction identifier and a replay cache mechanism that effectively protects the protocol against previously successful replay attacks.

*4) Message forgery protection:* The same unique transaction identifier, along with the computed TAG, acts as an additional defense layer to make sure of the authenticity of the sender and the message itself.

*5) Performance characterization:* Measuring the security-performance trade-off, providing experimental data that shows a robust security gain is achievable with a measured low overhead of (9.28% throughput decrease and 15.26% latency increase).

In Table III, a brief analysis comparison between the proposed solution and other RAFT implementations is presented.

TABLE III. PERFORMANCE ANALYSIS

| item | Original RAFT | RaBFT | Pathloss (Physical Layer) | Proposed solution |
|---|---|---|---|---|
| Primary Goal | Crash Fault Tolerance (CFT) and Log Consistency | Byzantine Fault Tolerance (BFT) | Authentication against Impersonation Attacks. | Mitigate Replay / Message Forgery attacks |
| Mitigation | None | Secret Sharing, Dynamic Committee Roles, Altered Election/Replication. | Pathloss Measurement as a Device Fingerprint. | Authenticated Encryption and Anti-Replay Cache. |
| Layer | Application | Application/Consensus (Protocol Redesign) | Physical | Transport |
| Overhead / Complexity | Simple, understandable logic. | High. Complete protocol redesign, complex multi-party computation. | Moderate. Requires initial calibration. | Low, minimal impact on core logic. |
| Weakness | vulnerable | High performance and complexity overhead. | Requires special hardware, susceptible to environmental changes. | Cache file size growth. |

The findings of this research confirm that the lack of message authentication is the biggest security weakness in the standard RAFT. Our solution provides a high-applicability framework for enhancing RAFT implementations that prioritize data integrity over raw speed and cannot tolerate the complexity of full Byzantine Fault Tolerance (BFT) protocols.

While highly effective against man-in-the-middle replay and forgery attacks, the solution's dependence on a per-message cache introduces a practical limitation concerning memory scalability and cache management over extended operational periods.

## XI. FUTURE WORK

Future research will focus on transitioning this modular solution into a production-ready framework by addressing its current limitations and expanding its current scope. The future research should include areas such as:

- Cache optimization: Implementing and evaluating time-to-live (TTL) and least recently used (LRU) eviction policies to ensure the replay cache maintains high efficiency and manages memory consumption dynamically.
- Secure Key Lifecycle Management: Developing a secure, dynamic process for periodic rotation of the master secret key to defend against long-term compromise of a server.
- Expanded attack areas validation: Broadening the scope of validation to test the solution against other network-level vulnerabilities, such as Denial-of-Service (DoS) attacks.